\documentclass[manuscript]{aastex}

\slugcomment{}

\shorttitle{Observing NEOs with JWST}
\shortauthors{Thomas et al.}

\begin{document}


\title{Observing Near-Earth Objects with the James Webb Space Telescope}


\author{Cristina A. Thomas\altaffilmark{1,2,3}}
\email{cristina.a.thomas@nasa.gov}

\author{Paul Abell\altaffilmark{4}}
\author{Julie Castillo-Rogez\altaffilmark{5}}
\author{Nicholas Moskovitz\altaffilmark{6}}
\author{Michael Mueller\altaffilmark{7,8}}
\author{Vishnu Reddy\altaffilmark{3}}
\author{Andrew Rivkin\altaffilmark{9}}
\author{Erin Ryan\altaffilmark{1,10}}
\and
\author{John Stansberry\altaffilmark{11}}


\altaffiltext{1}{NASA Goddard Space Flight Center, 8800 Greenbelt Rd., Greenbelt, MD 20771}
\altaffiltext{2}{NASA Postdoctoral Program, Oak Ridge Associated Universities}
\altaffiltext{3}{Planetary Science Institute, 1700 East Fort Lowell, Suite 106, Tucson, AZ}
\altaffiltext{4}{NASA Johnson Space Center, 2101 NASA Parkway, Houston, TX 77058}
\altaffiltext{5}{Jet Propulsion Laboratory, 4800 Oak Grove Dr., Pasadena, CA 91011}
\altaffiltext{6}{Lowell Observatory, 1400 West Mars Hill Rd., Flagstaff, AZ 86011}
\altaffiltext{7}{Rijksuniversiteit Groningen, Kapteyn Astronomical Institute, Postbus 800, 9700 AV Groningen, the Netherlands}
\altaffiltext{8}{SRON Netherlands Institute for Space Research, Astrophysics Research
Group, Postbus 800, 9700 AV Groningen, the Netherlands}
\altaffiltext{9}{JHU-Applied Physics Laboratory, 11100 Johns Hopkins Rd., Laurel, MD 20723}
\altaffiltext{10}{University of Maryland, Department of Astronomy, College Park, MD 20742}
\altaffiltext{11}{Space Telescope Science Institute, 3700 San Martin Drive, Baltimore, MD, 21218}


\begin{abstract}
The James Webb Space Telescope (JWST) has the potential to enhance our understanding of near-Earth objects (NEOs). We present results of investigations into the observability of NEOs given the nominal observing requirements of JWST on elongation (85-135 degrees) and non-sidereal rates ($<$30mas/s).
We find that approximately 75\% of NEOs can be observed in a given year. However, observers will need to wait for appropriate observing windows. We find that JWST can easily execute photometric observations of meter-sized NEOs which will enhance our understanding of the small NEO population. 

\end{abstract}





\section{Introduction}

The James Webb Space Telescope (JWST) is a 6.5 meter diameter space-based infrared observatory. The instrument suite includes NIRCam (Near-Infrared Camera), NIRISS (Near-Infrared Imager and Slitless Spectrograph), NIRSpec (Near-Infrared Spectrograph), and MIRI (Mid-Infrared Camera), which provide wavelength coverage from 0.6 to 28.5 microns with a variety of imaging and spectroscopic modes. JWST is scheduled for launch in late 2018 and is designed for a five year prime science mission with enough consumables onboard for ten years of operation. In the introduction to this special issue, \cite{Milam} discuss JWST's plan for operations and instrument capabilities with respect to Solar System observations. Additional information regarding JWST, the science goals, and mission implementation is discussed in \cite{Gardner06}.

Near-Earth Objects (NEOs) are asteroids and comets whose orbits bring them close to Earth's orbit. Due to multiple gravitational encounters with planetary bodies, the dynamical lifetime of an NEO is a few million years \citep{Bottke02}. Therefore, NEOs represent an ongoing flux of small bodies from elsewhere in the Solar System into near-Earth space. Once near Earth, these bodies become particularly important for several reasons: they are a population whose compositional diversity is directly related to their sources in other locations in the Solar System; their proximity to Earth enables observations of smaller objects than possible for other small body populations; they are the direct precursors to our meteorite collection; members of the population may pose an impact risk; and they are the closest bodies to the Earth-Moon system for direct exploration and potential resource utilization. 

The number of known NEOs has grown tremendously in the past two decades. According to the NASA Near Earth Object Program Office, there were $<$350 known NEOs in 1995 and more than 13,000 known by the end of 2015. This growth in discovery is largely due to several dedicated survey facilities and programs such as Pan-STARRS, the Catalina Sky Survey, and LINEAR. This number is expected to dramatically increase as additional surveys with larger aperture telescopes (such as the Large Synoptic Survey Telescope) begin operations. 

The James Webb Space Telescope (JWST) has the potential to enhance our understanding of the near-Earth Object population. The sensitivity of its instruments and the ability to reach wavelength regions that are unobservable from the ground will add tremendously to our understanding of surface composition, alteration, and physical properties.  

\section{JWST Observing Requirements}


JWST will orbit the Sun-Earth L2 point in a Lissajous orbit. This orbit guarantees continuous communication with Earth and requires the sunshield to be positioned such that the mirrors and instruments are protected from the heat of the Sun and Earth. The thermal requirements of the observatory and the corresponding requirements on the sunshield placement define the solar elongation window for JWST observations to be between 85 and 135 degrees. 

Non-sidereal tracking will be performed by moving the position of the guide star along a calculated ephemeris trajectory in the Fine Guidance Sensor (FGS), which is located in the same focal plane as the scientific instruments. The controlled motion of the guide star will hold the moving science target fixed in the instrument field of view. The operations plan for JWST allows for non-sidereal tracking of objects with rates $<$30 milliarcseconds per second (mas/s). This limit on non-sidereal rates is set such that pointing stability for non-sidereal moving targets is expected to be similar to sidereal targets \citep{Milam}.   

The most important question regarding observations of near-Earth objects with JWST is whether the telescope can track at the high non-sidereal rates that are characteristic of the NEO population while the objects are within the JWST field of regard. The rate of apparent motion of a Solar System object is dependent on the distance of that object from the observatory, the relative velocity of the object with respect to the observatory, and the orientation of the object's orbit relative to position of the observatory. During 2019, rates of all NEOs within the JWST elongation window (85-135 degrees) range from 0.001 to 3672 milliarcseconds/second (mas/s). To provide context for these NEO non-sideral rates, we calculated non-sidereal rates as viewed from JWST (within the elongation window) of subsets of three other small body populations during 2019-2020: Main Belt Asteroids (MBAs), comets, and Kuiper Belt Objects (KBOs) with centaurs. The subsets included over 100 objects for each group and were chosen to be representative of the overall population. Rates of motion for these three groups are 0.03-20 mas/s, 0.01-189 mas/s, and 0.003-2 mas/s, respectively. During this two year period, no MBAs or KBOs would be excluded from observations with JWST due to high rates of motion and a small number of comets would have their observation windows shortened, but not excluded, due to rates of motion exceeding 30 mas/s. We note that all calculations and discussions of non-sidereal rates of motion in this work use $d(RA)/dt\times cos(Dec)$ for right ascension (RA) and $d(Dec)/dt$ for declination (Dec).




\subsection{Quantifying the fraction of observable NEOs} 

We undertook two examinations of the NEO population with respect to the non-sidereal tracking limit in order to quantify the fraction of all known near-Earth objects ($\sim$12,000 at the end of 2014) that can be observed with the nominal observational constraints (elongation between 85-135 degrees and non-sidereal motion $<$30mas/s). 
To examine the non-sidereal motions of the entire near-Earth object population (known during October 2014) we used ephemerides from Lowell Observatory and JPL Horizons. 

We queried the Lowell ephemeris software for the positions of all known NEOs for the month of September 2014 at five day intervals. The resulting ephemerides assumed a ground based observatory. All known NEOs were queried and filtered based on the elongation window requirements and an orbital quality parameter which ensured that only objects with well constrained orbits were included in the sample. The total motion was calculated from the component RA and Dec coordinate motions. 
Given the nominal observing constraints, we determined that $\sim$14.0-14.5\% of the total NEO population can be observed on these September 2014 days.  
The total number of NEOs used in this calculation was 11,637 which was the total number of known NEOs as of 2014 October 1 according to the Near-Earth Object Program Office at the Jet Propulsion Laboratory. 
We use this investigation of NEOs using the Lowell ephemerides to estimate the percentage of NEOs that can be observed at any given time. However, since these objects are a relatively small fraction of the total population, this does not guarantee an accurate, unbiased sampling of the entire population. 

[Figure 1]

In order to examine the observability of the NEO population as a whole over a year (a period of time equivalent to a single observing cycle), we expanded our analysis by considering the ephemerides of all known NEOs from the expected position of JWST using the JPL Horizons ephemeris service. We calculated the positions of all known NEOs for each day from January 1, 2019 to January 1, 2020. The results were filtered based on the elongation and total motion requirements. Of the 11,467 near-Earth asteroids that were included in the investigation, only 2,754 had no observing window over the year long period. Approximately 75\% of the NEO population could be observed by JWST during 2019. The total number of NEOs in the 2019 Horizons investigation differs slightly from the Lowell sample despite the fact that the NEO population was defined at the same time. This discrepancy is due to the slightly different definitions used by the two methods. The Lowell sample included all known NEOs, but the Horizons sample only included those that were classified as Atens, Apollos, and Amors. Figure 1 shows that many objects have large enough observing windows that will enable ease of scheduling.

\subsection{Comparision with ExploreNEOs Warm Spitzer program}

The ExploreNEOs \citep{Trilling10,Trilling15} program was allocated 500 hours over two years (2009-2011) to determine albedos and diameters for 583 near-Earth objects using the warm mission capabilities of the IRAC (Infrared Array Camera, \citet{Fazio04}) instrument on the Spitzer Space Telescope \citep{Werner04}. ExploreNEOs was intended to be a statistical sampling of the entire NEO population and we assume the target list is representative of discovered NEOs. 
Due to the similarity in elongation requirements between JWST and Spitzer (82.5-120 degrees), we examine the ExploreNEOs target list to compare the non-sidereal tracking capabilities of JWST to the previous infrared space telescope.   
Spitzer is not at the L2 point as JWST will be and the Spitzer Observing Manual states that the observatory supports observations up to 1 arcsecond per second\footnote{Section 4.6 of Spitzer Space Telescope Observer's Manual}. The ExploreNEOs targets were checked for tracking rate before execution, but no objects were excluded due to a high tracking rate. The NEO with the fastest rate of motion at the time of observation was moving at 1955 arcseconds/hour or 543 mas/s.

We identified the rate of motion for each ExploreNEOs target at the time of observation using Spitzer Astronomical Observation Requests (AORs). At JWST's nominal tracking rate of $<$30 milliarcseconds per second, approximately 240 of the 583 Spitzer AORs or $\sim$40\% of ExploreNEOs objects would be observable. If the tracking rate was doubled to $<$60 mas/s, then $\sim$400 targets or $\sim$70\% of the ExploreNEOs objects would be accessible. In order for JWST to observe $\sim$500 objects ($\sim$85\%) from this sample the objects would need to be tracked with a limit of $<$100 mas/s. 

[Figure 2]

\section{Flux Calculations}\label{flux}

The other important variable in determining the observability of NEOs with JWST is the expected flux for members of the NEO population. In order to explore a range of observing circumstances, we examined six near-Earth objects. We identified observing windows between 1 January 2019 and 15 February 2023 (just before the end of the Horizons JWST ephemeris solution) for six NEOs with diameters from a few meters to tens of kilometers. We selected two large NEOs with very different albedos (Eros and Don Quixote), two objects with diameters of 1-2~km (2005 LR3 and 2001 SK162), and two very small NEOs with diameters of a few meters (2009 BD and 2011 MD). The four objects with diameters over a kilometer  represent the range of typical NEOs that have been observed by ground and space based resources. The smallest two objects were targeted by IRAC on Warm Spitzer, but were challenging observations and the calculated physical parameters are uncertain \citep{Mommert14b,Mommert14c}. We use these two objects as a proxy for the small NEO population that has not been well characterized by other telescopes and that JWST could help us better understand. The expected infrared flux for each object was calculated using the Near-Earth Asteroid Thermal Model (NEATM, \citet{Harris1998}) with estimated diameter, albedo, heliocentric distance, JWST-centric distance, and phase integral ($q\sim0.39$, e.g., \citet{Bowell1989,Trilling07,Lim11}). Observability is determined by comparing the flux in a nominal observing time (defined as 1000 seconds) to the sensitivity and saturation limits of the instruments. 

We limit our flux analysis to sensitivity measurements for the Near-Infrared Spectrograph (NIRSpec) and the Mid-Infrared Instrument (MIRI) since they will likely be the most heavily used for NEO and asteroid studies. NIRSpec is a near-infrared multi-object spectrograph with a 3.4' x 3.6' field of view, three spectral resolutions (R$\sim$100 for 0.7-5.0 $\mu$m in a single prism, R=1000 for 1.0-5.0 $\mu$m in three gratings, and R=2700 for 1.0-5.0 $\mu$m in three gratings), and three slit selection methods (array of micro-shutters, fixed slits, and a 3" x 3" Integral Field Unit). MIRI has direct imaging with a field of view of 1.25' x 1.88' for wavelengths 5-28.3 $\mu$m, low resolution spectroscopy (R$\sim$100) for wavelengths 5-10 $\mu$m, and a medium resolution spectroscopy (R$\sim$1000-3000) Integral Field Unit with a 3.5" x 3.5" field of view for wavelengths 5-27 $\mu$m. Figures 3 \& 4 show the calculated flux for each object at a single point in their JWST observing window. We examined the flux for the beginning and end of all observing windows described in the following text. For all objects except Eros, the figures show the expected flux on the brightest of those dates. We display the faintest calculated flux for Eros to show that saturation will be an issue for MIRI images even under the best circumstances. The fluxes included in figures 3 \& 4 are calculated for the following dates: Eros (7/15/19), Don Quixote (1/1/19), 2005 LR3 (3/18/20), 2001 SK162 (10/18/20), 2009 BD (7/29/22), and 2011 MD (2/2/23). 

{\em (433) Eros:} Eros is one of the largest NEO (average $D=17.5$) and has a relatively high geometric albedo (disk averaged $p_V=0.22$, \citet{Li04}). 
Eros meets the nominal observing criteria ($<$30 milliarcseconds per second within the solar elongation window) for 77 days from late April to mid-July 2019. Throughout this window, Eros can be observed by all NIRSpec gratings with the nominal 1000 second integration time. Eros can also be observed with all MIRI spectral gratings, but would saturate the detector using most of the MIRI imaging filters. 
We note that prior to the start of the 2019 observing window, Eros is close enough to JWST for the object to be spatially resolved with MIRI (pixel scale=0.11"). During this period Eros meets the elongation requirement, but the rate of motion is above the 30 mas/s limit. 

{\em (3552) Don Quixote:} Don Quixote is another of the largest NEOs ($D=18.4$ km) and has a very low geometric albedo ($p_V=0.03$, \citet{Mommert14a}). Don Quixote meets the nominal observing criteria over two observing windows for a total of 100 days during 2019.    
The heliocentric range(r) and the observer range ($\Delta$) of Don Quixote vary by $\sim$2 AU over the course of the two 2019 observing windows. Throughout the 100 observable days in 2019, Don Quixote can be observed by all spectral modes and filters in NIRSpec, but towards the end of the year as the object becomes farther from JWST and the Sun the calculated flux begins to fall below the required limits for the short wavelength end of the R=2700 NIRSpec grating. Don Quixote can be observed with all MIRI spectral gratings and filters with no saturation. 

{\em (198856) 2005 LR3:} (198856) is an asteroid (absolute magnitude, $H=16.9$) that is classified in the Q spectral class. Its spectral type suggests a relatively high geometric albedo (Q-type average $p_V=0.29$, \citet{Thomas11}) which we use to estimate a size of $D=1.03$ km. (198856) meets the nominal observing criteria for 54 days in 2020. During this observing window, 2005 LR3 can be observed by all spectral modes and filters in both NIRSpec and MIRI. 

{\em (162998) 2001 SK162:} (162998) is an NEO (absolute magnitude, $H=17.9$) that is classified as a D-type. We use the D-type average albedo of $p_V=0.02$ to estimate $D=2.47$ km. 2001 SK162 meets the nominal observing criteria over two observing windows for a total of 106 days during 2019. This object is observable with all NIRSpec filters, the NIRSpec R$\sim$100 prism, all MIRI filters, and most MIRI spectral modes.

{\em 2009 BD:} 2009 BD is an $H=28.1$ object which \citet{Mommert14b} estimate as either $D=2.9\pm0.3$ m with$p_V=0.85^{+0.2}_{-0.1}$ or $D=4\pm1$m with $p_V=0.45^{+0.2}_{-0.1}$ based on the upper-limit flux density from Spitzer observations. This NEO meets the nominal observing criteria for 91 days during 2022. 
We calculated the flux expected from 2009 BD using both sets of diameter and albedo estimates from \citet{Mommert14b}. 2009 BD can be observed by the wide broadband filters on NIRSpec and several of the MIRI filters. 

{\em 2011 MD:} 2011 MD is a $H=28$ object which \citet{Mommert14c} estimate as $D=6^{+4}_{-2}$ with $p_V=0.3^{+0.4}_{-0.2}$ based on Spitzer observations. This NEO enters the elongation window in February 2023 near the end of the Horizons JWST ephemeris solution. However, the object remains $\sim$40\% above the nominal tracking limit. We include this object since we use it as a proxy for the small NEO population. 2011 MD could be observed by the widest broadband filters on NIRSpec and several of the MIRI filters. 

These results suggest that both large and small NEOs can be observed by JWST, but an observer will likely have to wait for appropriate observing windows. Even the smallest objects, which are the most difficult to observe with other facilities, will have short observing windows where the object meets the elongation and non-sidereal motion criteria in addition to having the necessary flux for a successful observation. Our calculations suggest that with the nominal observing criteria, JWST can easily execute photometric observations of meter-size NEOs.
Unfortunately, these objects will not be bright enough to benefit from the full suite of observing capabilities of the NIRSpec and MIRI instruments. This will affect our ability to fully characterize the most unstudied objects and understand the variety of compositions present at the small size end of the NEO population. 

These calculations come with a couple of caveats. The chosen objects have spectral data and in some cases radiometrically derived albedos and diameters. We can use the spectral data to determine the taxonomic class and estimate the size for those without measured diameters. In many cases, this is far more information than an observer will have about an NEO prior to planning the JWST observation. The observer will have to estimate the albedo and diameter of the asteroid in order to estimate the flux which will lead to non-negligible propagated errors when calculating exposure times and signal-to-noise. 

One important caveat to these examinations of NEO rates of motion and fluxes is that NEO fluxes will have large variations over the course of an observability window. The brightness of an NEO is dependent on observer-centric distance, therefore NEOs tend to be brightest when they are fastest. This suggests that observers will have to find a balance between observing targets when they are bright and will have the highest signal to noise and the elongation and non-sidereal motion observing requirements. We can also expect NEOs that are not at the extreme ends of the known size distribution will have fewer restrictions on their observing window due to their fluxes.  

[Figure 3]

[Figure 4]

\section{Potential Science Cases}

{\em Understanding the Size Frequency Distribution of Small NEOs:} The size frequency distribution of small NEOs (D$<$100m) is not well understood \citep{Mainzer11,Trilling15}. In our flux calculations, we demonstrated that MIRI can easily obtain photometry of meter-sized NEOs that meet the elongation and motion requirements. These mid-infrared observations would enable calculation of albedo and diameters for an unexplored size regime of the NEO population. 

{\em Observations of Individual Objects:} Sometimes a single object can change our understanding of Solar System evolution. One example is the asteroid 2008 TC3 which was discovered shortly before impacting Earth and being recovered as the Almahata Sitta meteorite fall \citep{Jenn09}. The asteroid had a relatively flat reflectance spectrum, but the recovered meteorite has shown numerous disparate lithologies \citep{Bischoff10}. Prior to the discovery of the object, many asteroid researchers would not have expected such a large variation in the compositions of the meteorite fragments. Observations of other small, dark asteroids in near-Earth space may yield compositional results indicative of complicated formation histories. Unfortunately, objects with trajectories such as 2008 TC3, which was discovered shortly before impact, would not be observable by JWST since its position near Earth would be outside the elongation window. 

{\em Spacecraft Target Followup:} The capabilities of JWST enable scientific investigations in wavelength regions beyond those of ground-based observatories and spacecraft missions. Just as the Spitzer Space Telescope was used for observations of Eros and Itokawa, JWST can be used to further our understanding of future spacecraft targets such as (162173) 1999 JU3 and (101955) Bennu. One complication is that for objects in low delta-V orbits, there are long periods between observation windows. Hayabusa 2 target 1999 JU3 is observable in mid 2021, while OSIRIS-REx target Bennu is not observable until near the end of the nominal five year JWST mission in mid 2023.   

{\em Evolution of Dead or Dormant Comets:} The sensitivity of JWST will be extremely valuable for identifying outgassing on any dead or dormant comet candidates. The Spitzer Space Telescope identified outgassing from (3552) Don Quixote \citep{Mommert14a}, a large, dark NEO. Other comet candidates can be observed to place limits on any outgassing during the object's orbit. 

{\em Spectral Changes Between the Main Belt and Near-Earth Space:} \citet{Masiero15} demonstrated that the primitive Euphrosyne family could be an important contributor to the NEO population. However, the shape of the 3-$\mu$m band, a signature of hydration on the surface, does not look like those of the carbonaceous chondrites (e.g., \citet{Takir12}). Since we expect the NEO population to be the direct precursors to the meteorite collection, either the family is not producing meteorites or their spectra are changing between the Main Belt and near-Earth space. The higher surface temperatures experienced while in the NEO population could cause the loss of volatile species that are seen on the surface of primitive Main Belt asteroids. We can compare primitive NEOs with their Main Belt sources to identify any spectral changes that result from time in the high temperature near-Earth environment. 

\section{Conclusions}

JWST has the potential to enhance our understanding of the NEO population. The sensitivity of its instruments and the ability to reach wavelength regions that are unobservable from the ground will add tremendously to our understanding of surface composition, alteration, and physical properties. The design of the observatory defines the nominal observational requirements as non-sidereal motion $<$30mas/s with an elongation angle between 85-135 degrees.
We examined whether the NEO population can be observed by JWST given the high non-sidereal rates that are characteristic of the population. We calculated that approximately 75\% of the NEO population could be observed by JWST over the period of a year. 
Our analyses of calculated flux density suggest that both large and small NEOs can be observed by JWST, but an observer will have to wait for appropriate observing windows. We find that JWST can easily execute photometric observations of meter-sized NEOs, which are difficult to observe with other facilities. 
We have identified several potential NEO science questions that JWST will be able to address once the prime mission has begun.



\acknowledgments

C.A. Thomas was supported by an appointment to the NASA Postdoctoral Program at Goddard Space Flight Center, administrated by Oak Ridge Associated Universities through a contract with NASA.

\clearpage





\begin{figure}
\plotone{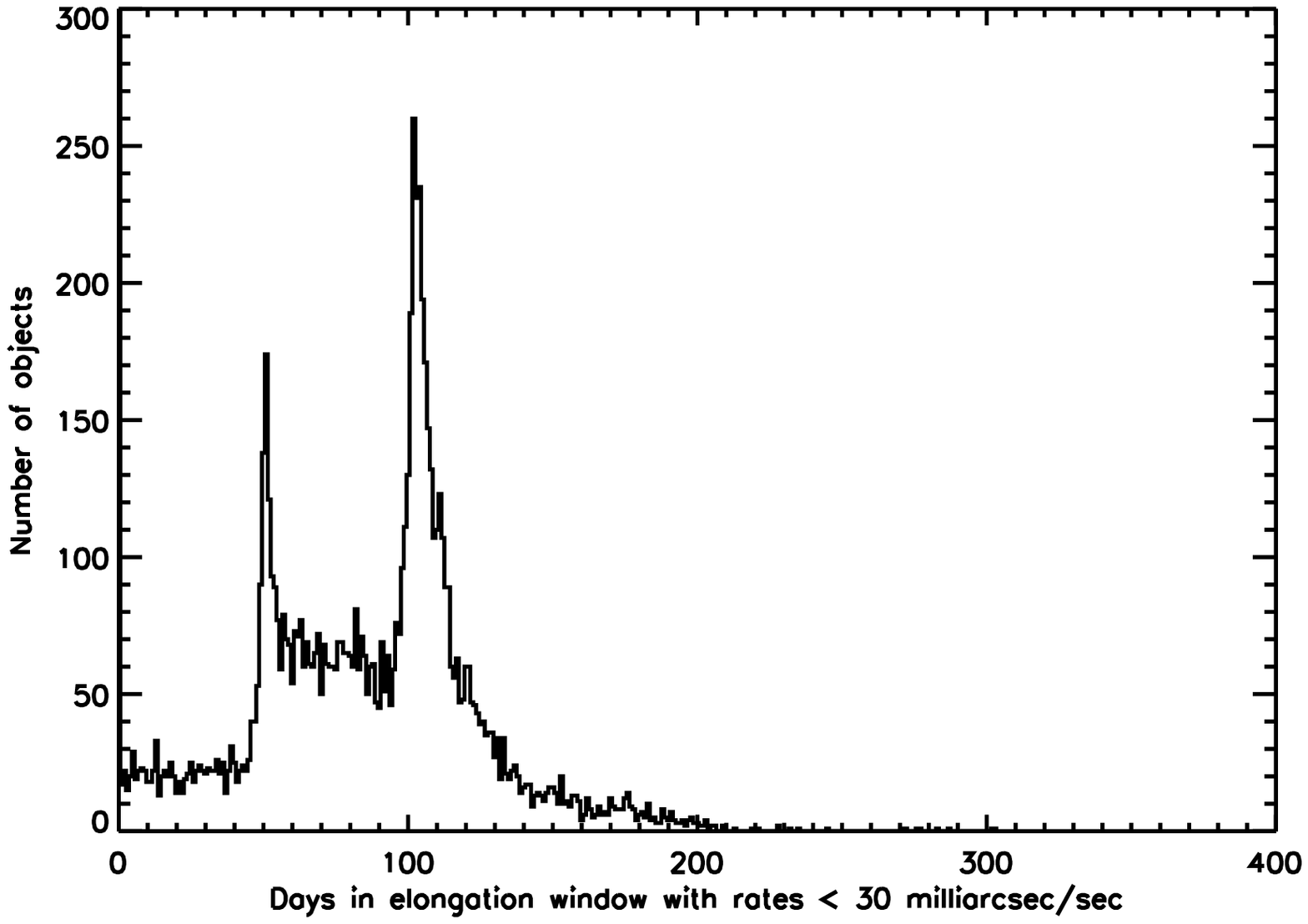}
\caption{Histogram showing the number of objects that can be observed by JWST with respect to the days in the elongation (85-135 degrees) and non-sidereal motion ($<$30mas/s ) requirements.}
\end{figure}
\clearpage


\begin{figure}
\includegraphics[scale=0.60,angle=270]{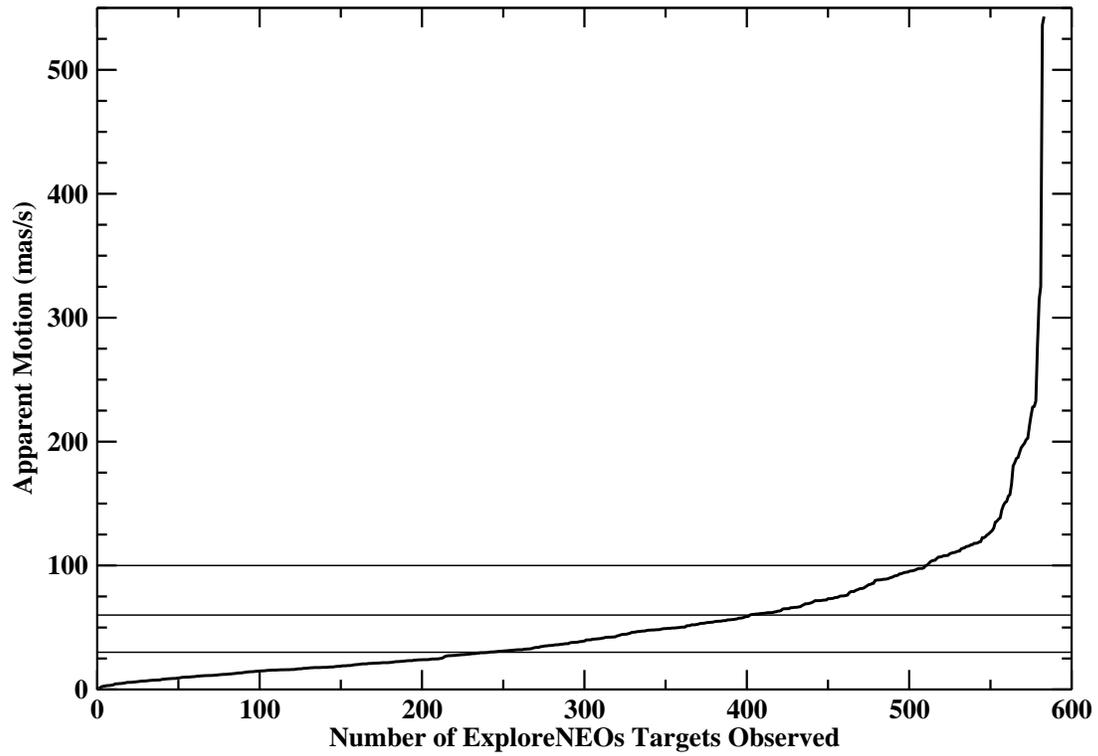}
\caption{Cumulative number of ExploreNEOs targets observable as a function of apparent non-sidereal motion in milliarcseconds per second. Horizontal lines highlight three apparent rates of motion: 30 mas/s (the nominal JWST tracking limit), 60 mas/s (double the nominal JWST limit), and 100 mas/s.}
\end{figure}

\clearpage

\begin{figure}
\epsscale{0.80}
\plotone{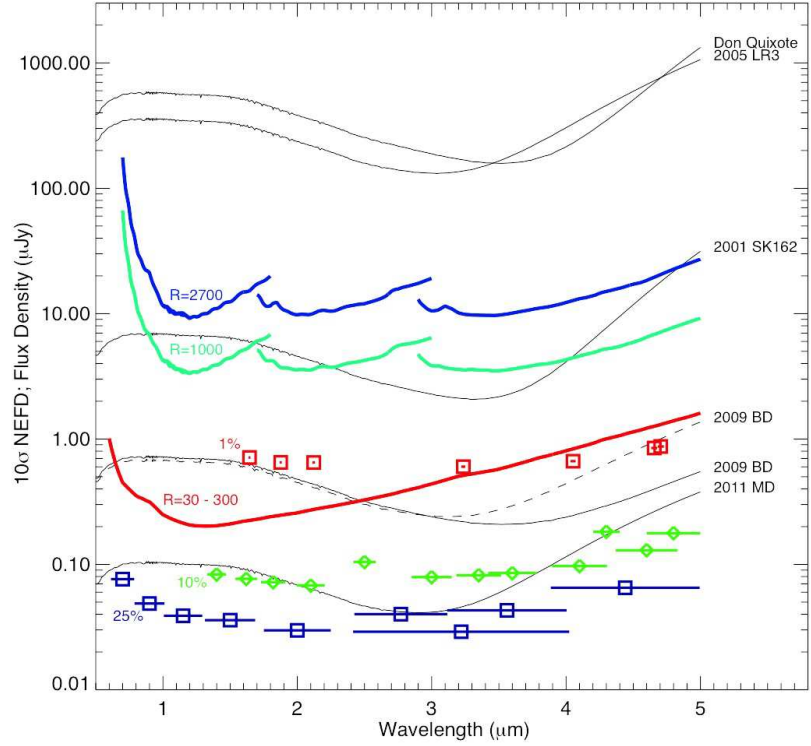}
\caption{Sensitivity and saturation limits for NEOs with the imaging and spectrographic modes of the NIRSpec instrument. The sensitivity limits of the NIRSpec filters are shown by the red, green, and blue horizontal lines with squares and diamonds at the center wavelength. The filters are divided into three groups which have bandpasses of approximately 25\%, 10\%, and 1\% of the filter wavelength. NIRSpec spectral sensitivity (noise-equivalent flux density, or NEFD) curves for the various spectral modes are shown in blue, green, and red with their respective spectral resolutions ($R$). We show calculated fluxes for five NEOs described in the text (Section \ref{flux}). The flux calculated for Eros is significantly higher than the scale of the figure. We include two flux calculations for 2009 BD to account for the two size and albedo estimates from \citet{Mommert14b}.
The integration time for all sensitivity values is 1000 seconds. }
\end{figure}

\begin{figure}
\epsscale{0.80}
\plotone{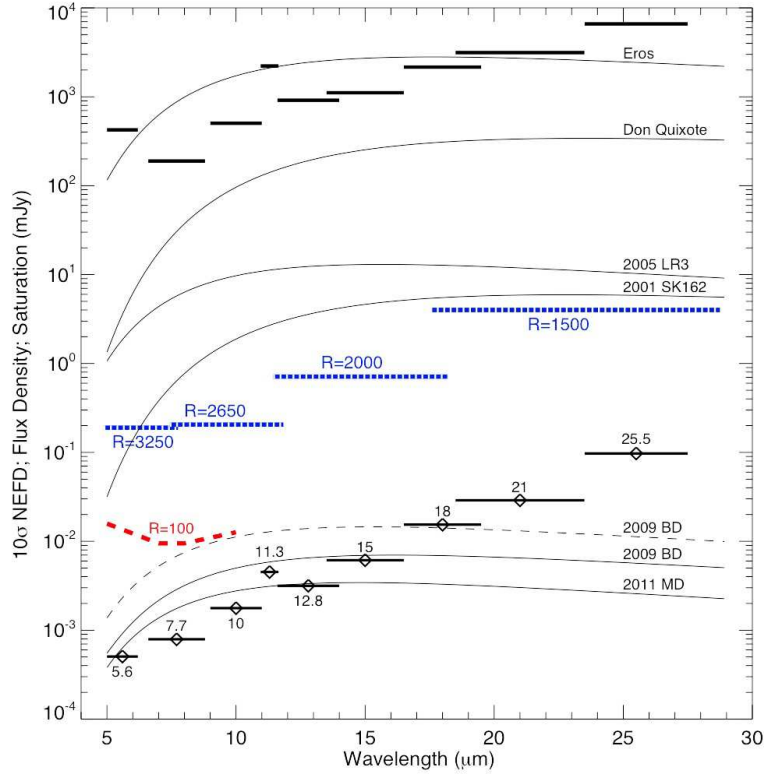}
\caption{Sensitivity and saturation limits for NEOs with the imaging and spectroscopic modes on the MIRI instrument. The sensitivity limits of the MIRI broadband filters are shown by the black horizontal lines with diamonds at the center wavelength. The corresponding saturation limits for these filters are shown by bold black lines of equivalent widths. The saturation limits are calculated for one frame with a subarray and, therefore, represent a hard upper limit on the observable point source flux.  
MIRI spectral sensitivity (noise-equivalent flux density, or NEFD) curves for the various spectral modes are shown by red and blue dashed lines with their respective spectral resolutions ($R$).  
We show calculated fluxes for the six NEOs as described in the text (Section \ref{flux}). We include two flux calculations for 2009 BD to account for the two size and albedo estimates from \citet{Mommert14b}.
The integration time for all sensitivity values is 1000 seconds.}
\end{figure}

\end{document}